\documentclass[a4paper,11pt]{article}
\usepackage[left=3.17cm,right=3.17cm,top=2.54cm,
headheight=0.5cm,headsep=0.54cm,bottom=2.54cm,footskip=0.79cm
]{geometry}

\usepackage[all]{xy}
\usepackage{amsmath,amsfonts,amssymb,amsthm,epsfig,amscd,comment,latexsym,psfrag}
\usepackage{nicefrac,xspace,tikz}

\usetikzlibrary{arrows}
\usetikzlibrary{decorations.markings}
\def\Z{\mathbb Z}

\def\C{\mathbb C}

\def\1{{\bf{1}}}

\usepackage[pdfstartview=FitH]{hyperref}

\def\footnoterule{\kern 1mm \hrule width 7cm \kern 2.2mm}%

 \newtheorem{thm}{Theorem}[section]
 
 \newtheorem{lem}[thm]{Lemma}

 \newtheorem{dfn}[thm]{Definition}

\newcommand{\bea}{\begin{eqnarray}}
\newcommand{\eea}{\end{eqnarray}}
\newcommand{\beaa}{\begin{eqnarray*}}
\newcommand{\eeaa}{\end{eqnarray*}}
\newcommand{\be}{\begin{equation}}
\newcommand{\ee}{\end{equation}}
\newcommand{\nn}{\nonumber}

\def\twoboxy{\begin{tikzpicture}
\draw (0,0)rectangle(0.5,0.25);
\draw (0.25,0) -- (0.25,0.25) [-];
\draw (0,0.25) -- (0.175,0.35) [-];
\draw [shift = {+(0.25,0)}](0,0.25) -- (0.175,0.35) [-];
\draw [shift = {+(0.5,0)}](0,0.25) -- (0.175,0.35) [-];
\draw [shift = {+(0.5,-0.250)}](0,0.25) -- (0.175,0.35) [-];
\draw (0.175,0.35) -- (0.675,0.35) [-];
\draw [shift = {+(0.25,0)}] (0.425,0.35) -- (0.425,0.1) [-];
\end{tikzpicture}}
\def\twoboxz{\begin{tikzpicture}
\draw (0,0)rectangle(0.25,0.5);
\draw (0,0.25) -- (0.25,0.25) [-];
\draw [shift = {+(0,0.25)}](0,0.25) -- (0.175,0.35) [-];
\draw [shift = {+(0.25,0.25)}](0,0.25) -- (0.175,0.35) [-];
\draw [shift = {+(0.25,0)}](0,0.25) -- (0.175,0.35) [-];
\draw [shift = {+(0.25,-0.250)}](0,0.25) -- (0.175,0.35) [-];
\draw (0.425,0.625) -- (0.425,0.1) [-];
\draw [shift = {+(0.175,0.35)}] (0,0.25) -- (0.25,0.25) [-];
\end{tikzpicture}}
\def\twoboxx{\begin{tikzpicture}
\draw (0,0)rectangle(0.25,0.25);

\draw (0,0.25) -- (0.35,0.45) [-];
\draw [shift = {+(0.25,0)}] (0,0.25) -- (0.35,0.45) [-];
\draw [shift = {+(0.25,-0.25)}] (0,0.25) -- (0.35,0.45) [-];
\draw [shift = {+(0.175,0.35)}](0,0) -- (0.25,0) [-];
\draw [shift = {+(0.25,-0.25)}][shift = {+(0.175,0.35)}](0,0) -- (0,0.25) [-];
\draw [shift = {+(0.35,0.45)}](0,0) -- (0.25,0) [-];
\draw [shift = {+(0.175,0.1)}] [shift = {+(0.25,-0.25)}][shift = {+(0.175,0.35)}](0,0) -- (0,0.25) [-];
\end{tikzpicture}}

\begin{document}

\title{3D Boson representation of affine Yangian of ${\mathfrak{gl}}(1)$ and 3D cut-and-join operators}
\author{Wang Na\dag\footnote{Corresponding author: wangna@henu.edu.cn },\ Zhang Can\dag\ Wu Ke\ddag\\
\dag\small School of Mathematics and Statistics, Henan University, Kaifeng, 475001, China\\
\ddag\small School of Mathematical Sciences, Capital Normal University, Beijing 100048, China}

\date{}
\maketitle

\begin{abstract}
We have constructed 3D Bosons. In this paper, we show the 3D Bosonic Fock space, which is isomorphic to the vector space of 3D Young diagrams as graded vector spaces. We use 3D Bosons to represent the generators of the affine Yangian of ${\mathfrak{gl}}(1)$ and define the 3D cut-and-join operators. Then we discuss the 3D Boson representation of $W$-operators in matrix models.
\end{abstract}

\section{Introduction}\label{sect1}
Schur functions and Jack polynomials defined on 2D Young diagrams are attractive research objects. Schur functions were used to determine irreducible characters of highest weight representations of the classical groups\cite{FH,Mac,weyl}. Recently they appear in mathematical physics, especially in integrable models. In \cite{MJD}, the group in the  Kyoto school uses Schur functions in a remarkable way to understand the KP and KdV hierarchies.  In \cite{NVT,PS}, Tsilevich and Su\l kowski, respectively, give the realization of the phase model in the algebra of Schur functions and build the relations between the $q$-boson model and Hall-Littlewood functions. In \cite{wang}, we build the relations between the statistical models, such as phase model, and KP hierarchy by using 2D Young diagrams and Schur functions. In \cite{WLZZ}, the authors show that the states in the $\beta$-deformed Hurwitz-Kontsevich matrix model can be represented as the Jack polynomials.

3-Jack polynomials defined on 3D Young diagrams (plane partitions) are a generalization of Schur functions and Jack polynomials. We have constructed 3-Jack polynomials in \cite{JHEP3Jack}. 3D Young diagrams arose naturally in crystal melting model\cite{ORV,NT}. 3D Young diagrams also have many applications in many fields of mathematics and physics, such as statistical models, number theory, representations of some algebras (Ding-Iohara-Miki algebras, affine Yangian, etc). 3-Jack polynomials behave the same with 3D Young diagrams when we consider the MacMahon representation of the affine Yangian of ${\mathfrak{gl}}(1)$.

We have constructed the 3D Bosons in \cite{JHEP3DBoson}. In this paper, we firstly construct the 3D Bosonic Fock space. As the 2D Bosonic Fock space is isomorphic to the space of 2D Young diagrams or the space of Schur functions defined on 2D Young diagrams, the 3D Bosonic Fock space is isomorphic to the space of 3D Young diagrams or the space of 3-Jack polynomials defined on 3D Young diagrams.

The MacMahon representation space of affine Yangian of ${\mathfrak{gl}}(1)$ is also the space of 3D Young diagrams\cite{Pro}. The main result of this paper is that we use 3D Bosons to realize the generators of the affine Yangian of ${\mathfrak{gl}}(1)$. In \cite{JHEP3DBoson}, we have obtained the representation of $W_{1+\infty}$ algebra, which have actions on 3D Young diagrams, by 3D Bosons, then we can see the relations between the operators in affine Yangian of ${\mathfrak{gl}}(1)$ and $W_{1+\infty}$ algebra since they both can be represented by 3D Bosons.

Use the operators $\psi_j$ in affine Yangian of ${\mathfrak{gl}}(1)$, we define 3D cut-and-join operators. 2D cut-and-join operators are operators commutative with each other, which have prominent applications in matrix models\cite{WLZZ}, the simplest one is
\[
\frac{1}{2}\sum_{n,m=1}^\infty \left((n+m)p_np_m\frac{\partial}{\partial p_{n+m}}+nmp_{n+m}\frac{\partial^2}{\partial p_n\partial p_m}\right).
\]
Schur functions are the common eigenstates of 2D cut-and-join operators. In this paper, we use 3D Bosons to represent 3D cut-and-join operators. The eigenstates are 3-Jack polynomials, we show that by some examples. The $W$-operators of the 3D generalizations of some matrix models can also be represented by 3D Bosons.

The paper is organized as follows. In section \ref{sect2}, we recall the $W_{1+\infty}$ algebra and the affine Yangian of ${\mathfrak{gl}}(1)$. In section \ref{sect3}, we recall 3D Bosons and give the 3D Bosonic Fock space.
 In section \ref{sect4},  we use 3D Bosons to represent the generators $\psi_3,\ e_0,\ f_0$ of affine Yangian of ${\mathfrak{gl}}(1)$.
In section \ref{sect5}, we construct 3D cut-and-join operators and use 3D Bosons to represent the operators in some matrix models.
\section{$W_{1+\infty}$ algebra and affine Yangian of ${\mathfrak{gl}}(1)$}\label{sect2}
In this section, we recall the $W_{1+\infty}$ algebra and the affine Yangian of ${\mathfrak{gl}}(1)$, which all have the representations on 3D Young diagrams.
\subsection{Affine Yangian of ${\mathfrak{gl}}(1)$}
Let $h_1,h_2$ and $h_3$ be three complex numbers satisfying $h_1+h_2+h_3=0$. Define
\beaa
\sigma_2 &=& h_1 h_2 + h_1 h_3 + h_2 h_3,\\
 \sigma_3 &=& h_1 h_2 h_3.
\eeaa
We associate $h_1,\ h_2,\ h_3$ to $y,\ x,\ z$-axis respectively.

The affine Yangian $\mathcal{Y}$ of ${\mathfrak{gl}}(1)$ is an associative algebra with generators $e_j, f_j$ and $\psi_j$, $j = 0, 1, \ldots$ and the following relations\cite{Pro,Tsy}
\begin{eqnarray}
&&\left[ \psi_j, \psi_k \right] = 0,\\
&&\left[ e_{j+3}, e_k \right] - 3 \left[ e_{j+2}, e_{k+1} \right] + 3\left[ e_{j+1}, e_{k+2} \right] - \left[ e_j, e_{k+3} \right]\nonumber \\
&& \quad + \sigma_2 \left[ e_{j+1}, e_k \right] - \sigma_2 \left[ e_j, e_{k+1} \right] - \sigma_3 \left\{ e_j, e_k \right\} =0,\label{yangian1}\\
&&\left[ f_{j+3}, f_k \right] - 3 \left[ f_{j+2}, f_{k+1} \right] + 3\left[ f_{j+1}, f_{k+2} \right] - \left[ f_j, f_{k+3} \right] \nonumber\\
&& \quad + \sigma_2 \left[ f_{j+1}, f_k \right] - \sigma_2 \left[ f_j, f_{k+1} \right] + \sigma_3 \left\{ f_j, f_k \right\} =0, \label{yangian2}\\
&&\left[ e_j, f_k \right] = \psi_{j+k},\label{yangian3}\\
&& \left[ \psi_{j+3}, e_k \right] - 3 \left[ \psi_{j+2}, e_{k+1} \right] + 3\left[ \psi_{j+1}, e_{k+2} \right] - \left[ \psi_j, e_{k+3} \right]\nonumber \\
&& \quad + \sigma_2 \left[ \psi_{j+1}, e_k \right] - \sigma_2 \left[ \psi_j, e_{k+1} \right] - \sigma_3 \left\{ \psi_j, e_k \right\} =0,\label{yangian4}\\
&& \left[ \psi_{j+3}, f_k \right] - 3 \left[ \psi_{j+2}, f_{k+1} \right] + 3\left[ \psi_{j+1}, f_{k+2} \right] - \left[ \psi_j, f_{k+3} \right] \nonumber\\
&& \quad + \sigma_2 \left[ \psi_{j+1}, f_k \right] - \sigma_2 \left[ \psi_j, f_{k+1} \right] + \sigma_3 \left\{ \psi_j, f_k \right\} =0,\label{yangian5}
\end{eqnarray}
together with boundary conditions
\begin{eqnarray}
&&\left[ \psi_0, e_j \right]  = 0, \left[ \psi_1, e_j \right] = 0,  \left[ \psi_2, e_j \right]  = 2 e_j ,\label{yangian6}\\
&&\left[ \psi_0, f_j \right]  = 0,  \left[ \psi_1, f_j \right]  = 0,  \left[ \psi_2, f_j \right]  = -2f_j ,\label{yangian7}
\end{eqnarray}
and a generalization of Serre relations
\begin{eqnarray}
&&\mathrm{Sym}_{(j_1,j_2,j_3)} \left[ e_{j_1}, \left[ e_{j_2}, e_{j_3+1} \right] \right]  = 0, \label{yangian8} \\
&&\mathrm{Sym}_{(j_1,j_2,j_3)} \left[ f_{j_1}, \left[ f_{j_2}, f_{j_3+1} \right] \right]  = 0,\label{yangian9}
\end{eqnarray}
where $\mathrm{Sym}$ is the complete symmetrization over all indicated indices which include $6$ terms.

The affine Yangian $\mathcal{Y}$ has a representation on 3D Young diagrams.
As in our paper \cite{3DFermionYangian}, we use the following notations. For a 3D Young diagram $\pi$, the notation $\Box\in \pi^+$ means that this box is not in $\pi$ and can be added to $\pi$. Here ``can be added'' means that when this box is added, it is still a 3D Young diagram. The notation $\Box\in \pi^-$ means that this box is in $\pi$ and can be removed from $\pi$. Here ``can be removed" means that when this box is removed, it is still a 3D Young diagram. For a box $\Box$, we let
\begin{equation}\label{epsilonbox}
h_\Box=h_1y_\Box+h_2x_\Box+h_3z_\Box,
\end{equation}
where $(x_\Box,y_\Box,z_\Box)$ is the coordinate of box $\Box$ in coordinate system $O-xyz$. Here we use the order $y_\Box,x_\Box,z_\Box$ to match that in paper \cite{Pro}.

Following \cite{Pro,Tsy}, we introduce the generating functions:
\begin{eqnarray}
e(u)&=&\sum_{j=0}^{\infty} \frac{e_j}{u^{j+1}},\nonumber\\
f(u)&=&\sum_{j=0}^{\infty} \frac{f_j}{u^{j+1}},\\
\psi(u)&=& 1 + \sigma_3 \sum_{j=0}^{\infty} \frac{\psi_j}{u^{j+1}},\nonumber
\end{eqnarray}
where $u$ is a parameter.
Introduce
\begin{equation}\label{psi0}
\psi_0(u)=\frac{u+\sigma_3\psi_0}{u}
\end{equation}
and
\begin{eqnarray} \label{dfnvarphi}
\varphi(u)=\frac{(u+h_1)(u+h_2)(u+h_3)}{(u-h_1)(u-h_2)(u-h_3)}.
\end{eqnarray}
For a 3D Young diagram $\pi$, define $\psi_\pi(u)$ by
\begin{eqnarray}\label{psipiu}
\psi_\pi(u)=\psi_0(u)\prod_{\Box\in\pi} \varphi(u-h_\Box).
\end{eqnarray}
In the following, we recall the representation of the affine Yangian on 3D Young diagrams as in paper \cite{Pro} by making a slight change. The representation of affine Yangian on 3D Young diagrams is given by
\begin{eqnarray}
\psi(u)|\pi\rangle&=&\psi_\pi(u)|\pi\rangle,\\
e(u)|\pi\rangle&=&\sum_{\Box\in \pi^+}\frac{E(\pi\rightarrow\pi+\Box)}{u-h_\Box}|\pi+\Box\rangle,\label{eupi}\\
f(u)|\pi\rangle&=&\sum_{\Box\in \pi^-}\frac{F(\pi\rightarrow\pi-\Box)}{u-h_\Box}|\pi-\Box\rangle\label{fupi}
\end{eqnarray}
where $|\pi\rangle$ means the state characterized by the 3D Young diagram $\pi$ and the coefficients
\begin{equation}\label{efpi}
E(\pi \rightarrow \pi+\square)=-F(\pi+\square \rightarrow \pi)=\sqrt{\frac{1}{\sigma_3} \operatorname{res}_{u \rightarrow h_{\square}} \psi_\pi(u)}.
\end{equation}
Specially,
\bea
&&\psi_1|\pi\rangle=0,\ \psi_2|\pi\rangle=2|\pi||\pi\rangle,\ \psi_3|\pi\rangle=\sum_{\Box\in\pi}(6h_\Box+2\psi_0\sigma_3)|\pi\rangle,\nn\\
&& \psi_4|\pi\rangle=\sum_{\Box\in\pi}(12h_\Box^2-2\sigma_2+6h_\Box\psi_0\sigma_3)|\pi\rangle,\label{psipi}
\eea
where $|\pi|$ is the box number of $\pi$.

In the following of this paper, we treat $E(\pi \rightarrow \pi+\square)|\pi+\square\rangle$ as one element and still denote it by $|\pi+\square\rangle$, then 3D Young diagrams depend on the box growth process. 3-Jack polynomials $J_\pi$ behave the same as $\pi$ exactly.

\subsection{$W_{1+\infty}$ algebra}
For any 3D Young diagram, we cut it into slices by the plane $z=j$, every slice is a 2D Young diagram. Then 3D Young diagrams can be treated as a series of 2D Young diagrams. When we consider the 3D Young diagrams which have at most $N$ layers in $z$-axis direction, we suppose $\psi_0=-\frac{N}{h_1h_2}$. Let $a_{j,n}$ be the 2D Bosons associated to the 2D Young diagrams on the slice $z=j$ of 3D Young diagrams with the relation
\be\label{ajnakmcom}
[a_{j,n},a_{k,m}]=-\frac{1}{h_1h_2}\delta_{j,k}n\delta_{n+m,0}.
\ee
The operators $V_{j,n}$ of the $W_{1+\infty}$ algebra can be represented by $a_{j,n}, \ j=1,2,\cdots,N$, which corresponds to that a 3D Young diagram can be represented by a series of 2D Young diagrams. This means that the operators $V_{j,n}$ have the representation on the space of 3D Young diagrams.

The fields
\be
J_j(z)=\sum_{n\in\Z}a_{j,n}z^{-n-1}\ \ \text{and}\ \ V_j(z)=\sum_{n\in\Z}V_{j,n}z^{-j-n}.
\ee
Let\cite{JHEP3DBoson}
\bea
V_1(z)&=&J_1(z)+J_2(z)+\cdots+J_N(z),\label{v1zU}\\
V_2(z)&=&-\frac{h_1h_2}{2}\sum_{j=1}^N :J_j(z)J_j(z):-\frac{h_1h_2\alpha_0}{2}\sum_{j=1}^N(N+1-2j) J_j^\prime(z),\label{v2zU}\\
V_3(z)&=&\frac{1}{3}h_1^2h_2^2\sum_{j=1}^N:J_1(z)^3:-\frac{1}{2}\alpha_0h_1^2h_2^2\sum_{j<k}J_jJ_k^\prime(z)+\frac{1}{2}\alpha_0h_1^2h_2^2\sum_{j<k}J_j^\prime J_k(z)\nn\\
&&+\frac{1}{2}\alpha_0h_1^2h_2^2\sum_{j=1}^N(N+1-2j)J_j^\prime J_j(z)\nn\\
&&-\alpha_0^2h_1^2h_2^2\sum_{j=1}^N\left(\frac{(j-1)(N-j)}{2}-\frac{(N-1)(N-j)}{12}\right)J_j^{\prime\prime}(z),\label{v3zU}
\eea
and\cite{3KP}
\beaa
V_4(z)&=&-h_1^3h_2^3\left(\frac{1}{4}\sum_{j=1}^NJ_jJ_jJ_jJ_j(z)-\frac{\alpha_0}{2}\sum_{j<k}J_jJ_jJ_k'(z)
-{\alpha_0}\sum_{j<k<l}(l-3)J_jJ_k'J_l(z)\right.\nn\\
&&+\frac{\alpha_0}{2}\sum_{j<k}J_jJ_k'J_k(z)+\frac{\alpha_0}{2}\sum_{j<k}J_j'J_jJ_k(z)+\frac{\alpha_0}{2}\sum_{j<k}J_j'J_kJ_k(z)\nn\\
&&-{\alpha_0}\sum_{j<k<l}(l-3)J_j'J_kJ_l(z)+\frac{\alpha_0}{2}\sum_{j=1}^N(N+1-2j)J_j'J_jJ_j(z)\\
&&-\frac{\alpha_0^2}{4}\sum_{j<k}(N+1-2k)J_jJ_k''(z)-\frac{\alpha_0^2}{4}\sum_{j<k}(N+1-2j)J_j''J_k(z)\\
&&\left.-\alpha_0^2\sum_{j<k}(k-j)J_j'J_k'(z)+\frac{3}{20h_1h_2}\sum_{j=1}^NJ_j'J_j'(z)-\frac{1}{10h_1h_2}\sum_{j=1}^NJ_j''J_j(z)+\cdots\right).
\eeaa

Specially, when $N=1, h_1=1,h_2=-1$, 3D Young diagrams become 2D Young diagrams and 3-Jack polynomials become Schur functions. We denote $V_j(z)$ in this special case by $V_j^{2DS}(z)$. Then
\bea
{V}_1^{2DS}(z)&=&J_1(z),\\
{V}_2^{2DS}(z)&=&\frac{1}{2}:J_1(z)^2:,\\
{V}_3^{2DS}(z)&=&\frac{1}{3}:J_1(z)^3:,\\
{V}_4^{2DS}(z)&=&\frac{1}{4}:J_1(z)^4:-\frac{3}{20}:J_1'(z)^2:+\frac{1}{10}:J_1''(z)J_1(z):\label{v42DSz}
\eea
with the OPE
\[
J_1(z)J_(w)\sim \frac{1}{(z-w)^2}.
\]
This special case $V_j^{2DS}(z)$ of the $W_{1+\infty}$ algebra have representation on the space of 2D Young diagrams or the Schur functions defined on 2D Young diagrams. The general case $V_j(z)$ of the $W_{1+\infty}$ algebra have representation on the space of 3D Young diagrams or the 3-Jack polynomials defined on 3D Young diagrams.

The OPEs of $V_j(z)V_k(w)$ can be found in \cite{JHEP3DBoson,3KP}. From the OPEs, the relations $[V_{j,m},V_{k,n}]$ can be obtained. We list the first few of them:
\begin{align*}
[V_{1,m},V_{1,n}]=&-\frac{N}{h_1h_2}m\delta_{m+n,0}=\psi_0 m\delta_{m+n,0},\\
[V_{1,m},V_{2,n}]=&m V_{1,m+n},\\
[ V_{2,m},V_{2,n} ]=&(m-n)V_{2,m+n}-(\psi_0\sigma_2+\psi_0^3\sigma_3^2) \frac{m^{3}-m }{12}\delta _{m+n,0},\\
[ V_{1,m},V_{3,n} ]=&2mV_{2,m+n}, \\
[ V_{2,m},V_{3,n} ] =&-\frac{1}{6}(\sigma_2+\sigma_3^2\psi_0^2)(m^3-m)V_{1,m+n}+(2m-n)V_{3,m+n}.
\end{align*}

\section{3D Bosons and 3D Bosonic Fock space}\label{sect3}
We have shown the method to construct the 3D Bosons in \cite{JHEP3DBoson}. In this section, we recall 3D Bosons and give the 3D Bosonic Fock space.

We denote 3D Bosons by $b_{n,j}$. When 3D Young diagrams have at most $N$ layers in the $z$-axis direction, $j$ in $b_{n,j}$ equals $1,2,\cdots, N$, which means that $b_{n,j}=0$ when $j>N$. Specially, when $N=1$, $b_{n,j}$ becomes $b_{n,1}$, which is the normal 2D Bosons. This special case corresponds to that 2D Young diagrams can be treated as the 3D Young diagrams which have one layer in $z$-axis direction. 3D Bosons $b_{n,j}$ can be represented by 2D Bosons $a_{j,n}$:
\beaa
b_{n,1}&=&\sum_{j=1}^Na_{j,n},\\
b_{n,2}&=&-h_1h_2(1-\frac{1}{N})\sum_{j=1}^N\sum_{k+l=n}:a_{j,k}a_{j,l}:+\frac{2h_1h_2}{N}\sum_{j<k}\sum_{m+l=n}:a_{j,m}a_{k,l}:\\
&&+h_3\sum_{j=1}^N(N+1-2j)(-n-1)a_{j,n},\\
b_{n,3}&=&6h_1^2h_2^2\left(-\sum_{j<k<l}\sum_{m+p+q=n}:a_{j,m}a_{k,p}a_{l,q}:-\alpha_0\sum_{j<k}\sum_{m+q=n}(j-1)(-m-1):a_{j,m}a_{k,q}:\right.\\
&&-\alpha_0\sum_{j<k}\sum_{m+q=n}(k-2)(-q-1):a_{j,m}a_{k,q}:-\frac{\alpha_0^2}{2}\sum_{j=1}^N(j-1)(j-2)(-n-1)(-n-2)a_{j,n}\\
&&+\frac{N-2}{N}\sum_{j=1}^N\sum_{k<l}\sum_{m+p+q=n}:a_{j,m}a_{k,p}a_{l,q}:+\frac{(N-2)\alpha_0}{N}\sum_{j,k=1}^N\sum_{m+q=n}(k-1)(-q-1):a_{j,m}a_{k,q}:\\
&&-\frac{(N-1)(N-2)}{3N^2}\sum_{j,k,l=1}^N\sum_{m+p+q=n}:a_{j,m}a_{k,p}a_{l,q}:+\frac{(N-2)\alpha_0^2}{2}\sum_{j=1}^N(j-1)(-n-1)(-n-2)a_{j,n}\\
&&+\frac{(N-2)\alpha_0}{2}\sum_{j<k}\sum_{m+q=n}(-m-q-2):a_{j,m}a_{k,q}:-\frac{(N-1)(N-2)\alpha_0^2}{12}\sum_{j=1}^N(-n-1)(-n-2)a_{j,n}\\
&&\left.-\frac{(N-1)(N-2)\alpha_0}{2N}\sum_{j,k=1}^N\sum_{m+q=n}(-m-1):a_{j,m}a_{k,q}:\right).
\eeaa
The relations between 3D Bosons can be obtained from the OPEs $B_j(z)B_k(w)$\cite{JHEP3DBoson}, we list some of them:
\begin{align}
[b_{m,1},b_{n,1}]=&\ \psi_0 m\delta_{m+n,0},\\
[b_{m,1},b_{n,j\geq 2}]=&\ 0,\\
[b_{m,2},b_{n,2}]=&\ 2(m-n)b_{m+n,2}-2(1+\sigma_2\psi_0+\psi_0^3\sigma_3^2)\frac{m^3-1}{6}\delta_{m+n,0},\\
[b_{m,2},b_{n,3}]=&\ 2(2m-n)b_{m+n,3}.
\end{align}

 Denote the vacuum state associated to 2D Bosons $a_{j,n}$ by $|0\rangle_j$, and Define
\[
|0\rangle=\sum_{j=1}^N|0\rangle_j.
\]
We know that $b_{n,j}|0\rangle=0$ for any $n>0$, and $b_{n,j}|0\rangle=0$ for any $n\leq 0, j>-n$. Denote the algebra generated by 3D Bosons $b_{n,j}$ by $\mathfrak{B}$. Define the 3D Bosonic Fock space by
\be
\mathfrak{B}\cdot |0\rangle:=\{a|0\rangle\ |\ a\in\mathfrak{B}\}.
\ee
Then 3D Bosonic Fock space has a basis
\be\{b_{-n_1,j_1}b_{-n_2,j_2}\cdots b_{-n_r,j_r}|0\rangle\ |\ 0<n_1\leq n_2\leq\cdots\leq n_r, j_i\leq n_i\}.\ee
Note that
\[
b_{-n,1}b_{-m,j\geq 2}|0\rangle=b_{-m,j\geq 2}b_{-n,1}|0\rangle,
\]
which is the same with that for 2D, but
$b_{-n,i\geq 2}b_{-m,j\geq 2}|0\rangle$ and $b_{-m,j\geq 2}b_{-n,i\geq 2}|0\rangle$ may do not equal to each other.
For example,
\[
b_{-2,2}b_{-3,2}|0\rangle=b_{-3,2}b_{-2,2}|0\rangle+2b_{-5,2}|0\rangle.
\]

Define the degree of $b_{-n,j}$ by $n$, and the degree of $|0\rangle$ by zero. Then 3D Bosonic Fock space is a graded space. The degree zero part is $\C|0\rangle$. The degree one part is $\C b_{-1,1}|0\rangle$. A basis of degree $2$ part is $\{b_{-1,1}^2|0\rangle,b_{-2,1}|0\rangle,b_{-2,2}|0\rangle\}$. The 3D Bosonic Fock space is isomorphic to the space of 3D Young diagrams, it is also isomorphic to the space of 3-Jack polynomials.
Let \be
P_{n,k}=b_{-n,k}|0\rangle
\ee
as in \cite{JHEP3DBoson}, and we denote
\[
b_{-n_1,j_1}b_{-n_2,j_2}\cdots b_{-n_r,j_r}|0\rangle
\]
by $P_{n_1,j_1}P_{n_2,j_2}\cdots P_{n_r,j_r}$, which explains the variables in 3-Jack polynomials. Note that for any 3-Jack polynomial $J_\pi$,
only $P_{n,1}J_\pi$ equals the normal multiplication of $P_{n,1}$ and $J_\pi$, $P_{n,j\geq 2}J_\pi$ equals the action $b_{-n,j\geq 2}\cdot J_\pi$. We give an example. Let
\beaa
a_{j,-n}|0\rangle_j=p_{j,n}|0\rangle_j,\ \ a_{j,n}|0\rangle_j=-\frac{1}{h_1h_2}\frac{\partial}{\partial p_{j,n}}|0\rangle_j,\ \ n>0,
\eeaa
where $p_{j,n}$ is the normal power sum on the slice $z=j$ of 3D Young diagrams. We know that\cite{3DHurwitz}
\beaa
P_{2,1}&=&\sum_{j=1}^N p_{j,2},\\
P_{2,2}&=& -h_1h_2\sum_{j=1}^N p_{j,1}^2+\frac{h_1h_2}{N}(\sum_{j=1}^N p_{j,1})^2-\sum_{j=1}^N(N-2j+1)h_3p_{j,2}.
\eeaa
It can be calculated that $P_{2,1}P_{2,2}$ equals the normal multiplication
\[
\sum_{j=1}^N p_{j,2}\left(-h_1h_2\sum_{j=1}^N p_{j,1}^2+\frac{h_1h_2}{N}(\sum_{j=1}^N p_{j,1})^2-\sum_{j=1}^N(N-2j+1)h_3p_{j,2}\right),
\]
but $P_{2,2}P_{2,2}$ does not equal
\[
\left(-h_1h_2\sum_{j=1}^N p_{j,1}^2+\frac{h_1h_2}{N}(\sum_{j=1}^N p_{j,1})^2-\sum_{j=1}^N(N-2j+1)h_3p_{j,2}\right)^2.
\]
\section{3D Boson representation of affine Yangian of ${\mathfrak{gl}}(1)$}\label{sect4}
In this section, we use 3D Bosons to represent the affine Yangian of ${\mathfrak{gl}}(1)$. From the relations in  the affine Yangian of ${\mathfrak{gl}}(1)$, we only need to represent the operators $\psi_3,\ e_0,\ f_0$.
\begin{thm}
The affine Yangian of ${\mathfrak{gl}}(1)$ can be represented by the 3D Bosons $b_{n,j}$ in the following way:
\bea
e_0&=&b_{-1,1},\label{e03dboson}\\
f_0&=&-b_{1,1},\label{f03dboson}\\
\psi_3&=& -\frac{1}{2}b_{0,3}+\frac{3}{\psi_0}\sum_{n>0}(b_{-n,1}b_{n,2}+b_{-n,2}b_{n,1})\nn\\
&&+\frac{3}{\psi_0^2}\sum_{n,m>0}(b_{-n,1}b_{-m,1}b_{n+m,1}+b_{-n-m,1}b_{n,1}b_{m,1})\nn\\
&&+3\sigma_3\sum_{n>0}nb_{-n,1}b_{n,1}-\sigma_3\psi_0\frac{\psi_2}{2},\label{psi33dboson}
\eea
with
\be
\psi_2=b_{0,2}+\frac{2}{\psi_0}\sum_{n>0}b_{-n,1}b_{n,1}.\label{psi23dboson}
\ee
\end{thm}

From \cite{WBCW}, we know that the operators of the affine Yangian of ${\mathfrak{gl}}(1)$ can be represented by a series of 2D Bosons $a_{j,n}$:
\bea
\psi_2&=&-2h_1h_2\sum_{j=1}^N\sum_{n>0}a_{j,-n}a_{j,n},\\
\psi_3 &=&3h_1^2h_2^2\sum_{j=1}^N\sum_{n,m>0}(a_{j,-n-m}a_{j,n}a_{j,m}
+a_{j,-n}a_{j,-m}a_{j,n+m})\nn\\
&&+6\sigma_3\sum_{j<k}\sum_{n>0}na_{j,-n}a_{k,n}+(-4N+6j-3)\sigma_3\sum_{j=1}^N\sum_{n>0}a_{j,-n}a_{j,n}\nn\\
&&+3\sigma_3\sum_{j=1}^N\sum_{n>0}na_{j,-n}a_{j,n},
\eea
and
\bea
e_0=\sum_{j=1}^N a_{j,-1},\ f_0=-\sum_{j=1}^N a_{j,1}.
\eea
It is clear that the relations (\ref{e03dboson}) and (\ref{f03dboson}) hold. The relations (\ref{psi33dboson}) and (\ref{psi23dboson}) can be proved by direct calculation, since $b_{n,j}$ can be represented by this series of 2D Bosons $a_{j,n}$. In the following, we obtain (\ref{psi33dboson}) and (\ref{psi23dboson}) by a simpler way. We need two lemmas.
\begin{lem}
The representation of $V_{3,0}$ by a series of 2D Bosons $a_{j,n}$ is
\bea
V_{3,0}&=&h_1^2h_2^2\sum_{j=1}^N\sum_{n,m>0}(a_{j,-n-m}a_{j,n}a_{j,m}
+a_{j,-n}a_{j,-m}a_{j,n+m})\nn\\
&&-\sigma_3\sum_{j<k}\sum_{n>0}\left(na_{k,-n}a_{j,n}-na_{j,-n}a_{k,n}\right)\nn\\
&&-\sigma_3\sum_{j=1}^N\sum_{n>0}(N-2j+1)a_{j,-n}a_{j,n}.
\eea
\end{lem}
This result can be obtained from (\ref{v3zU}).
\begin{lem}
The representation of $V_{3,0}$ by 3D Bosons $b_{n,j}$ is
\bea
V_{3,0}&=&-\frac{1}{6}b_{0,3}+\frac{1}{\psi_0}\sum_{n>0}(b_{-n,1}b_{n,2}+b_{-n,2}b_{n,1})\nn\\
&&+\frac{1}{\psi_0^2}\sum_{n,m>0}(b_{-n,1}b_{-m,1}b_{n+m,1}+b_{-n-m,1}b_{n,1}b_{m,1}).
\eea
\end{lem}
This result is obtained from
\be
V_3(z)=-\frac{1}{6}B_3(z)+\frac{1}{\psi_0}B_1B_2(z)+\frac{1}{3\psi_0^2}B_1B_1B_1(z).
\ee
Note that this relation is slightly different from that in \cite{JHEP3DBoson} since $V_3(z)$ here equals $V_3(z)$ in \cite{JHEP3DBoson} multiplied by $-1$.

The proof of (\ref{psi33dboson}).
\beaa
\psi_3-3V_{3,0}&=&3\sigma_3\sum_{j<k}\sum_{n>0}\left(na_{k,-n}a_{j,n}+na_{j,-n}a_{k,n}\right)\\
&&+3\sigma_3\sum_{j=1}^N\sum_{n>0}na_{j,-n}a_{j,n}-N\sigma_3\sum_{j=1}^N\sum_{n>0}a_{j,-n}a_{j,n}\\
&=&3\sigma_3\sum_{n>0}b_{-n,1}b_{n,1}-\sigma_3\psi_0\frac{\psi_2}{2}.
\eeaa

The proof of (\ref{psi23dboson}). It holds since
\be
\psi_2=2V_{2,0},
\ee
and
\be
V_{2}(z)=\frac{1}{2}B_2(z)+\frac{1}{2\psi_0}B_1B_1(z).
\ee
This relation is obtained in \cite{JHEP3DBoson}.

We give some examples about the eigenstates of $\psi_2$ and $\psi_3$ by (\ref{psi23dboson}) and (\ref{psi33dboson}). The eigenstates of $\psi_2$: For 3D Young diagram of one box,
\[
\psi_2b_{-1,1}|0\rangle=\frac{2}{\psi_0}b_{-1,1}b_{1,1}b_{-1,1}|0\rangle=2b_{-1,1}|0\rangle.
\]
For 3D Young diagram of two boxes,
\beaa
\psi_{2}b_{-1,1}^{2} |0\rangle &=&\frac{2}{\psi_{0 }} b_{-1,1}b_{1,1}b_{-1,1}^{2}|0\rangle=4b_{-1,1}^{2}|0\rangle,\\
\psi_{2}b_{-2,1}|0\rangle &=&\frac{2}{\psi_{0 }} b_{-2,1}b_{2,1}b_{-2,1}|0\rangle=4b_{-2,1}|0\rangle,\\
\psi_{2}b_{-2,2}|0\rangle &=& b_{0,2}b_{-2,2}|0\rangle=4b_{-2,2}|0\rangle.
\eeaa
For 3D Young diagram of three boxes,
\beaa
\psi_{2}b_{-1,1}^{3} |0\rangle &=&\frac{2}{\psi_{0 }} b_{-1,1}b_{1,1}b_{-1,1}^{3}|0\rangle=6b_{-1,1}^{3}|0\rangle,\\
\psi_{2}b_{-1,1}b_{-2,1}|0\rangle &=&\frac{2}{\psi_{0 }} b_{-1,1}b_{1,1}b_{-1,1}b_{-2,1}|0\rangle+\frac{2}{\psi_{0 }} b_{-2,1}b_{2,1}b_{-1,1}b_{-2,1}|0\rangle\\
&=&6b_{-1,1}b_{-2,1}|0\rangle,\\
\psi_{2}b_{-1,1}b_{-2,2}|0\rangle &=&\frac{2}{\psi_{0 }} b_{-1,1}b_{1,1}b_{-1,1}b_{-2,2}|0\rangle+ b_{0,2}b_{-1,1}b_{-2,2}|0\rangle\\
&=&6b_{-1,1}b_{-2,2}|0\rangle,\\
\psi_{2}b_{-3,1}|0\rangle &=&\frac{2}{\psi_{0 }} b_{-3,1}b_{3,1}b_{-3,1}|0\rangle=6b_{-3,1}|0\rangle,\\
\psi_{2}b_{-3,2}|0\rangle &=&b_{0,2}b_{-3,2}|0\rangle=6b_{-3,2}|0\rangle,\\
\psi_{2}b_{-3,3}|0\rangle &=&b_{0,2}b_{-3,3}|0\rangle=6b_{-3,3}|0\rangle.
\eeaa
The eigenstates of $\psi_3$: For 3D Young diagram of one box,
\[
\psi_3b_{-1,1}|0\rangle=3\sigma_3b_{-1,1}b_{1,1}b_{-1,1}|0\rangle-\sigma_3\psi_0\frac{\psi_2}{2}b_{-1,1}|0\rangle
=2\sigma_3\psi_0b_{-1,1}|0\rangle.
\]
For 3D Young diagram of two boxes,
\beaa
\psi_{3}b_{-1,1}^{2} |0\rangle &=&\frac{3}{\psi_{0 }^{2}} b_{-2,1}b_{1,1}b_{1,1}b_{-1,1}^{2}|0\rangle+3\sigma _{3}b_{-1,1}b_{1,1}b_{-1,1}^{2}|0\rangle-\sigma _{3}\psi _{0} \frac{\psi _{2}}{2}b_{-1,1}^{2}|0\rangle\\
&=&6b_{-2,1}|0\rangle+3\sigma _{3} \cdot 2\psi _{0}b_{-1,1}^{2}|0\rangle- 2\sigma _{3} \psi _{0}b_{-1,1}^{2}|0\rangle\\
&=&6b_{-2,1}|0\rangle+4\sigma _{3} \psi _{0}b_{-1,1}^{2}|0\rangle,\\
\psi_{3}b_{-2,1}|0\rangle &=&\frac{3}{\psi_{0 }} b_{-2,2}b_{2,1}b_{-2,1}|0\rangle+\frac{3}{\psi_{0 }^{2}}b_{-1,1}b_{-1,1}b_{2,1}b_{-2,1}|0\rangle\\
&&+3\sigma _{3} 2b_{-2,1}b_{2,1}b_{-2,1}|0\rangle-\sigma _{3}\psi _{0}\frac{\psi_{2 }}{2}b_{-2,1}|0\rangle\\
&=&6 b_{-2,2}|0\rangle+\frac{6}{\psi_{0 }}b_{-1,1}^{2}|0\rangle+12\sigma _{3}\psi _{0} b_{-2,1}|0\rangle-2\sigma _{3}\psi _{0}b_{-2,1}|0\rangle\\
&=&6 b_{-2,2}|0\rangle+\frac{6}{\psi_{0 }}b_{-1,1}^{2}|0\rangle+10\sigma _{3}\psi _{0} b_{-2,1}|0\rangle,\\
\psi_{3}b_{-2,2}|0\rangle &=&\frac{3}{\psi_{0 }} b_{-2,1}b_{2,2}b_{-2,2}|0\rangle-\sigma _{3}\psi _{0}\frac{\psi_{2 }}{2}b_{-2,2}|0\rangle\\
&=&-\frac{6}{\psi_{0 }}(1+\sigma _{2}\psi _{0}+\sigma _{3}^{2} \psi _{0}^{3} ) b_{-2,1}|0\rangle-2\sigma _{3}\psi _{0}b_{-2,2}|0\rangle,
\eeaa
then we have
\beaa
&&\psi _{3} \left((1+h_{2} h_{3})\frac{1}{\psi _{0} } b_{-1,1}^{2}|0\rangle+(1+h_{2} h_{3}\psi _{0})h_{1}b_{-2,1}|0\rangle+b_{-2,2}|0\rangle\right)\\
&=&(6h_{1}+4\sigma _{3}\psi _{0})\left((1+h_{2} h_{3}\psi _{0})\frac{1}{\psi _{0}} b_{-1,1}^{2}|0\rangle+(1+h_{2} h_{3}\psi _{0})h_{1}b_{-2,1}|0\rangle+b_{-2,2}|0\rangle\right),
\eeaa
which means that
\beaa
&&\tilde{J}_{\twoboxy}=\frac{1}{(h_1-h_2)(h_1-h_3)}\left((1+h_2h_3\psi_0)\frac{1}{\psi _{0}}P_1^2+(1+h_2h_3\psi_0)h_1P_{2,1}+P_{2,2}\right)\\
&=&\frac{1}{(h_1-h_2)(h_1-h_3)}\left((1+h_{2} h_{3}\psi _{0})\frac{1}{\psi _{0}} b_{-1,1}^{2}|0\rangle+(1+h_{2} h_{3}\psi _{0})h_{1}b_{-2,1}|0\rangle+b_{-2,2}|0\rangle\right)\\
\eeaa
is the eigenstate of $\psi_3$ with eigenvalue $(6h_{1}+4\sigma _{3}\psi _{0})$. Similarly, $\tilde{J}_{\twoboxx}$ and $\tilde{J}_{\twoboxz}$ are the eigenstates with eigenvalues $(6h_{2}+4\sigma _{3}\psi _{0})$ and $(6h_{3}+4\sigma _{3}\psi _{0})$ respectively.
\section{3D cut-and-join operators and some matrix models}\label{sect5}
In this section, we construct 3D cut-and-join operators and use 3D Bosons to represent the operators in some matrix models.
\subsection{2D/3D cut-and-join operators}
Since Schur functions defined on 2D Young diagrams are the eigenstates of the 2D cut-and-join operators, in this subsection, we consider the special case $N=1,h_1=-1,h_2=-1$ of $W_{1+\infty}$ algebra and affine Yangian of $\mathfrak{gl}(1)$, where $N$ is the at most layers of 3D Young diagrams in $z$-axis direction. Under this special case, 3D Young diagrams become 2D Young diagrams, and 3-Jack polynomials become Schur functions.

We denote $a_{1,n}$ by $a_n$, clearly
\[
[a_m,a_n]=m\delta_{m+n,0}
\]
specially, which have an representation on Schur functions by
\[
a_{-n}=p_n,\ \ \ a_n=n\frac{\partial}{\partial p_n}
\]
for $n>0$. The first two examples of the cut-and-join operators in terms of the power sums $p_n$ are\cite{cutandjoin}
\beaa
W_1&=&\sum_{n=1}^\infty np_n\frac{\partial}{\partial p_n},\\
W_2&=&\frac{1}{2}\sum_{n,m=1}^\infty \left((n+m)p_np_m\frac{\partial}{\partial p_{n+m}}+nmp_{n+m}\frac{\partial^2}{\partial p_n\partial p_m}\right).
\eeaa

We use the notation $\psi_j^{2DS}$ to denote $\psi_j$ in the special case $N=1,h_1=-1,h_2=-1$, and the same for other operators in the affine Yangian of $\mathfrak{gl}(1)$ and the operators in the $W_{1+\infty}$ algebra. Clearly,
\bea
&&\frac{1}{2}\psi_2^{2DS}=V_{2,0}^{2DS}=W_1,\\
&&\frac{1}{3}\psi_3^{2DS}=V_{3,0}^{2DS}=W_2.\label{psiv302ds}
\eea
Then, we can treat $W_{j=1,2}$ is the reduction of $V_{j+1,0}$ or $\frac{1}{j+1}\psi_{j+1}$ from 3D to 2D. We will define the 3D cut-and-join operators from $\psi_j$ for two reasons. The first one is that it can be checked that 3-Jack polynomials are not the eigenstates of $V_{3,0}$, while they are the eigenstates of all $\psi_j$.
For the second reason, we consider the special case $N=1$ and $h_1, h_2$ arbitrary. We denote $\psi_j$ in this special case by $\psi_j^{2D}$ and similarly for other operators. Then
\bea
V_{3,0}^{2D}&=&h_1^2h_2^2\sum_{n,m>0}(a_{-n-m}a_{n}a_{m}
+a_{-n}a_{-m}a_{n+m})\nn\\
\psi_3^{2D} &=&3h_1^2h_2^2\sum_{n,m>0}(a_{-n-m}a_{n}a_{m}
+a_{-n}a_{-m}a_{n+m})\nn\\
&&+\sigma_3\sum_{n>0}(3n-1)a_{-n}a_{n}
\eea
with $[a_n,a_m]=-\frac{1}{h_1h_2}n\delta_{n+m,0}$.
The eigenstates of $\psi_3^{2D}$ are Jack polynomials, and in the deformed Hurwitz-Kontsevich model, the $W$-operator
\beaa
&&\frac{1}{2}\sum_{k,l=1}^\infty \left(klp_{k+l}\frac{\partial}{\partial p_{k}}\frac{\partial}{\partial p_{l}}-h_1h_2(k+l)p_kp_l\frac{\partial}{\partial p_{k+l}}\right)\nn\\
&+&\frac{1}{2}\sum_{k=1}^\infty((h_1+h_2)(k-1)+2\psi_0\sqrt{\beta}N) k p_k\frac{\partial}{\partial p_{k}}
\eeaa
equals
\[
\frac{1}{6}\psi_3^{2D}+\frac{1}{2}(\psi_0\sqrt{\beta} N-\frac{1}{3}\psi_0\sigma_3)\psi_2^{2D}
\]
with $a_{-n}=p_n,\ a_n=-\frac{1}{h_1h_2}\frac{\partial}{\partial p_n} $ for $n>0$.

For this two reasons, we define 3D cut-and-join operators as follows.
\begin{dfn}
The 3D cut-and-join operators $W_n^{3D}$ are defined by
\be
W_n^{3D}=\frac{1}{n+1}\psi_{n+1}.
\ee
\end{dfn}
For example,
\beaa
W_1^{3D}&=&\frac{1}{2}\psi_2=\frac{1}{2}b_{0,2}+\frac{1}{\psi_0}\sum_{n>0}b_{-n,1}b_{n,1},\\
W_2^{3D}&=&\frac{1}{3}\psi_3=-\frac{1}{6}b_{0,3}+\frac{1}{\psi_0}\sum_{n>0}(b_{-n,1}b_{n,2}+b_{-n,2}b_{n,1})\nn\\
&&+\frac{1}{\psi_0^2}\sum_{n,m>0}(b_{-n,1}b_{-m,1}b_{n+m,1}+b_{-n-m,1}b_{n,1}b_{m,1})\nn\\
&&+\sigma_3\sum_{n>0}nb_{-n,1}b_{n,1}-\frac{1}{6}\sigma_3\psi_0b_{0,2}-\frac{1}{3}\sigma_3\sum_{n>0}b_{-n,1}b_{n,1}.
\eeaa
From the properties of $\psi_j$, we know that every two 3D cut-and-join operators are commutative, and the 3-Jack polynomials are their eigenstates.
\subsection{Some matrix models}
In this subsection, we use 3D Bosons to represent the $W$-operators in matrix models.

We recall the partition function hierarchy first. The Hurwitz-Kontsevich model \cite{Shakirov2009}
\begin{eqnarray}
Z_{0}\{p\}=\int_{\tilde{N}\times \tilde{N}} \sqrt{{\rm det}\left(\frac{{\rm sinh}(\frac{\phi\otimes I-I\otimes\phi}{2})}
{\frac{\phi\otimes I-I\otimes\phi}{2}} \right)}d\phi e^{-\frac{1}{2t}{\rm Tr}\phi^2
-\frac{\tilde{N}}{2}{\rm Tr}\phi-\frac{1}{6}t\tilde{N}^3+\frac{1}{24}t\tilde{N}+{\rm Tr}(e^{\phi}\psi)},
\end{eqnarray}
where  $\psi$ is an $\tilde{N}\times \tilde{N}$ matrix and the time variables $p_k={\rm Tr}\psi^k$. Here $\tilde{N}$ has no relations with $N$ before.
It is generated by the exponent of the Hurwitz operator $\hat{W}_0$ acting on the function
$e^{p_1/e^{t\tilde{N}}}$\cite{WLZZ},
\begin{eqnarray}\label{HKPF}
Z_{0}\{p\}=e^{t\hat{W}_0}\cdot e^{p_1/e^{t\tilde{N}}}
=\sum_{\lambda}e^{tc_{\lambda}}S_\lambda\{p_k=e^{-t\tilde{N}}\delta_{k,1}\} S_{\lambda}\{p\},
\end{eqnarray}
where $t$ is a deformation
parameter and $S_\lambda$ are Schur functions, the Hurwitz operator $\hat{W}_0$ is given by
\begin{eqnarray}
\hat{W}_0=\frac{1}{2}\sum_{k,l=1}^{\infty}\big((k+l)p_{k}p_l
\frac{\partial}{\partial p_{k+l}}+klp_{k+l}\frac{\partial}{\partial p_k}\frac{\partial}{\partial p_l}\big)
+\tilde{N}\sum_{k=1}^{\infty}kp_{k}\frac{\partial}{\partial p_{k}},
\end{eqnarray}
and $c_{\lambda}=\sum_{(i,j)\in \lambda}(\tilde{N}-i+j)$.

Introduce
\begin{equation}
E_{1}=[\hat{W}_0, p_1],\ \ \ \hat{W}_{-1}=[\hat{W}_0, E_{1}],
\end{equation}
the operators
\[
\hat{W}_{-n}=\frac{1}{(n-1)!}\underbrace{[\hat{W}_{-1}, [\hat{W}_{-1}, \cdots [\hat{W}_{-1} }_{n-1}, E_{1}]\ldots ]],\ \ n\geq 2\]
give the partition function hierarchies which include the Gaussian hermitian one-matrix model, $\tilde{N}\times\tilde{N}$ complex matrix model\cite{WLZZ}.

In the following, we use 3D Bosons to represent the 3D generations of these $W$-operators.
The 3D Hurwitz-Kontsevich model \cite{3DHurwitz}
\begin{equation}
{ Z}_{0}^{3D}\{p\}
=e^{t\hat{{W}}_0^{3D}}\cdot e^{ \frac{P_{1,1}}{\psi_0e^{tM}}},
\end{equation}
where
\begin{eqnarray}
\hat{{W}}_0^{3D}&=&\frac{1}{2}\sum_{i=1}^N\sum_{k,l=1}^{\infty}\big(klp_{i,k+l}\frac{\partial}{\partial p_{i,k}}\frac{\partial}{\partial p_{i,l}}-h_1h_2(k+l)p_{i,k}p_{i,l}
\frac{\partial}{\partial p_{i,k+l}}\big)\nn\\
&&+(h_1+h_2)\sum_{i_1<i_2}\sum_{k>0}k^2p_{i_1,k}\frac{\partial}{\partial p_{i_2,k}}\nonumber\\
&&+\frac{1}{2}\sum_{j=1}^N\sum_{k=1}^{\infty}\big((h_1+h_2)(k-2N+2j-1)+2\psi_0\sqrt{\beta} \tilde{N}\big)kp_{j,k}\frac{\partial}{\partial p_{j,k}}.\label{w0Jkoperator}
\end{eqnarray}
This operator $\hat{{W}}_0^{3D}$ can be represented by 3D Bosons
\bea
\hat{{W}}_0^{3D}&=&-\frac{1}{12}b_{0,3}+\frac{1}{2\psi_0}\sum_{n>0}(b_{-n,1}b_{n,2}+b_{-n,2}b_{n,1})\nn\\
&&+\frac{1}{2\psi_0^2}\sum_{n,m>0}(b_{-n,1}b_{-m,1}b_{n+m,1}+b_{-n-m,1}b_{n,1}b_{m,1})+\frac{1}{2}\sigma_3\sum_{n>0}nb_{-n,1}b_{n,1}\nn\\
&&+\frac{1}{2}\left(\psi_0\sqrt{\beta}\tilde{N}-\frac{1}{2}\psi_0\sigma_3\right)\left(b_{0,2}+\frac{2}{\psi_0}\sum_{n>0}b_{-n,1}b_{n,1}\right).
\eea
The 3D Boson representation of other operators can be obtained from this equation, for example,
\bea
E_1^{3D}&=&[W_0^{3D},b_{-1,1}]\nn\\
&=&\frac{1}{2}b_{-1,2}+\frac{1}{\psi_0}\sum_{n>0}b_{-n-1}b_{n,1}+\psi_0\sqrt{{\beta}}\tilde{N}b_{-1,1}.
\eea
 \section*{Data availability statement}
The data that support the findings of this study are available from the corresponding author upon reasonable request.

\section*{Declaration of interest statement}
The authors declare that we have no known competing financial interests or personal relationships that could have appeared to influence the work reported in this paper.

\section*{Acknowledgements}
This research is supported by the National Natural Science Foundation
of China under Grant No. 12101184 and No. 11871350, and supported by the Key Scientific Research Project in Colleges and Universities of Henan Province No. 22B110003.

\end{document}